\documentclass[referee]{raa}            

\usepackage{graphicx,times}             
\usepackage{natbib}
\usepackage{amssymb,amsmath}
\bibpunct{(}{)}{;}{a}{}{,}

\usepackage[a4paper=true,dvipdfm=true,pagebackref=true]{hyperref}
\hypersetup{colorlinks = true, linkcolor = green, anchorcolor = red, citecolor = blue, filecolor = red, pagecolor = red, urlcolor = red}

\begin{document}

   \title{Gas infall in the massive star formation core G192.16-3.84
}

   \volnopage{Vol.0 (20xx) No.0, 000--000}      
   \setcounter{page}{1}          

   \author{Mengyao Tang
      \inst{1}
   \and Sheng-Li Qin
      \inst{1}
   \and Tie Liu
      \inst{2,3}
   \and Yuefang Wu
      \inst{4}
   }

   \institute{Department of Astronomy, Yunnan University, and Key Laboratory of Astroparticle Physics of Yunnan Province, Kunming, 650091, China, mengyao\_tang@yeah.net, slqin@bao.ac.cn\\
        \and
             Korea Astronomy and Space Science Institute, 776 Daedeokdaero, Yuseong-gu, Daejeon 34055, Republic of Korea, liu@kasi.re.kr\\
        \and
             East Asian Observatory, 660 N. A'ohoku Place, Hilo, HI 96720, USA\\
        \and
            Department of Astronomy, Peking University, 100871, Beijing China, ywu@pku.edu.cn\\
\vs\no
   {\small Received~~20xx month day; accepted~~20xx~~month day}}

\abstract{ Previous observations have revealed an accretion disk and outflow motion in high-mass star-forming region G192.16-3.84.
While collapse have not been reported before.
We present here molecular line and continuum observations toward massive core G192.16-3.84 with the Submillimeter Array.
C$^{18}$O(2-1) and HCO$^{+}$(3-2) lines show pronounced blue profiles, indicating gas infalling in this region.
This is the first time that the infall motion has been reported in G192.16-3.84 core.
Two-layer model fitting gave infall velocities of 2.0$\pm$0.2 and 2.8$\pm$0.1 km s$^{-1}$. Assuming that the cloud core follows a power-law density profile ($\rho$$\propto$$r^{1.5}$), the corresponding mass infall rates are (4.7$\pm$1.7)$\times10^{-3}$ and (6.6$\pm$2.1)$\times10^{-3}$ M$_{\sun}$ yr$^{-1}$ for C$^{18}$O(2-1) and HCO$^{+}$(3-2), respectively.
The derived infall rates are in agreement with the turbulent core model and those in other high-mass star-forming regions, suggesting that high accretion rate is a general requirement to form a massive star.
}

   \authorrunning{M. Tang, S.-L. Qin, T. Liu, \& Y. Wu }            
   \titlerunning{Gas infall in star formation core G192.16-3.84}  

   \maketitle

%
%
\section{Introduction}
\label{sect:intro}
Current observational evidences suggest that the low-mass star
formation typically starts with a collapsing core inside a
molecular cloud. Then, the protostellar objects increase their
mass by gas accretion. Meanwhile, it is also accompanied by
outflows and accretion disk. Collapse, accretion disk and outflow, therefore,
are key elements in low mass star formation. However, the physical
conditions and dynamical processes of high-mass star formation are
still not well understood, due to observational difficulties caused by their short lifetimes and large distances.
Outflows are often found in the high-mass
star-forming regions \citep{Wu04,Qin08,Qiu12}. Only a handful of disks
in high-mass young stellar objects, however, have been detected
\citep{Zhang98,Shepherd01,Jiang05,Patel05,Sridharan05,Sanche14}.

G192.16-3.84 (hereafter G192.16) is a massive
protostellar system located at a distance of 1.52$\pm$0.08 kpc
\citep{Shiozaki11}. The luminosity of $\sim$3$\times$10$^{3}$
$L_{\sun}$ implies the presence of an early B star with a mass of
8 to 10 M$_{\sun}$ in this region \citep{Shepherd96,Shepherd98}.
Rich H$_{\rm 2}$O masers \citep{Shepherd04,Imai06,Shiozaki11}, UC
H$_{\rm II}$ region \citep{Hughes93,Shepherd99}, bipolar CO outflows
\citep{Shepherd98,Hauyu13}, rotational motions \citep{Hauyu13} and a solar system-size accretion disk
\citep{Shepherd01} have been observed in G192.16 region,
suggesting that massive star is forming in this region. However,
collapse of G192.16 core has not been reported before.

In this paper, we present
Submillimeter Array (SMA)\footnote{The Submillimeter Array is a
joint project between the Smithsonian Astrophysical Observatory
and the Academia Sinica Institute of Astronomy and Astrophysics
and is funded by the Smithsonian Institution and the Academia
Sinica.} observations of 230 GHz, 265 GHz, 345 GHz band data towards G192.16, showing collapsing motions in this region.

\section{Data}
\label{sect:Data}

All observational data used in our work are taken from the Submillimeter Array (SMA) archive.
The 230 GHz, 265 GHz, and 345 GHz observations were performed with SMA in
August 2005, December 2006, and December 2011, respectively.
The 230 GHz data cover CO(2-1), $^{13}$CO(2-1), C$^{18}$O(2-1), and SO(6$_{5}$-5$_{4}$) lines with an uniform spectral resolution of 0.8125 MHz. HCO$^{+}$(3-2) and HCN(3-2) transitions were observed in 265 GHz  band with hybrid high-spectral resolution. The 265 GHz data have different spectral resolutions in different windows. We resample the 265 GHz band data to uniform resolution of 0.8125 MHz. The 345 GHz data have spectral resolution of 0.8125 MHz and include CO(3-2) and SO(8$_{8}$-7$_{7}$) lines.
Other observational informations such as phase tracking center, bandpass calibrators, gain calibrators, and flux calibrators are listed in Table 1. Data reduction and imaging were made in  MIRIAD \citep{Sault95}.
The continuum images were made from line free channels.
Self-calibration on the continuum data were made to remove residual errors,
and then the gain solutions were applied to line data. The synthesized beam
sizes of continuum are summarized in Table 1.

\begin{table}[!h]
\footnotesize
\begin{center}
\caption[]{ SMA Observations}\label{Tab:publ-works}
 \begin{tabular}{clclclclc}
  \hline\noalign{\smallskip}
    \hline\noalign{\smallskip}
Phase Tracking Center & Band       & N$_{ant}$$^{a}$  & & Calibrator & & Beam Size                \\
\cline{4-6}
(R.A, Decl.)          &                           &            &Bandpass &Gain &Flux &$\arcsec$$\times$$\arcsec$($\degr$) \\
  \hline\noalign{\smallskip}
(5$^{h}$58$^{m}$13$^{s}$.899, 16$\degr$31$\arcmin$59$\arcsec$.997) &230 GHZ  &8  &3C454.3 &0530+135,0510+180 &Uranus    &3.74$\arcsec$$\times$3.04$\arcsec$(-86$\degr$)  \\
(5$^{h}$58$^{m}$13$^{s}$.530, 16$\degr$31$\arcmin$58$\arcsec$.300) &265 GHz  &8  &3C273   &0528+134,0507+179 &Titan     &0.86$\arcsec$$\times$0.86$\arcsec$(85$\degr$)  \\
(5$^{h}$58$^{m}$13$^{s}$.549, 16$\degr$31$\arcmin$58$\arcsec$.300) &345 GHz  &8  &3C84,Uranus &0530+135,0730-116 &Titan &1.80$\arcsec$$\times$1.59$\arcsec$(-58$\degr$)  \\
  \noalign{\smallskip}\hline
  $^{a}$ Number of Antennas \\
\end{tabular}
\end{center}
\end{table}

\section{Results}
\label{sect:results}

\subsection{Continuum}
Figure 1 presents the continuum flux density maps in both color-scale and contours.
From Figure 1, one can see that the continuum images at the three wavebands
show compact source structure and are unresolved.

Two dimension (2D) Gaussian fitting was made to the compact core. The peak
position of the continuum is R.A.(J2000) = 5$^{h}$58$^{m}$13$^{s}$.547, Decl.(J2000) = 16$\degr$31$\arcmin$58$\arcsec$.206, which is consistent with that of previous
continuum observations and UC H$_{\rm II}$ region \citep{Shepherd98,Shepherd99,Shepherd01,Shiozaki11,Hauyu13}.
The deconvolved size, peak flux density, total flux from Gaussian fitting are given in Table 2.

Continuum at our observed wavebands contains free-free emission ($S_{\nu}$ $\propto$ $\nu^{-0.1}$). Based on measured total flux of 1.5 mJy at the 3.6 cm band \citep{Shepherd99}, we estimate that the free-free continuum emission are 1.07 mJy and 1.03 mJy at the 230 GHz and 345 GHz bands. Comparing with total flux of the continuum at 230 and 345 GHz  (0.270 to 0.769 Jy), the free-free continuum emission is negligible.

To derive physical parameters , we performed spectral energy distribution (SED) fitting
based on the data from our observations and the previous data at different
wavelengths \citep{Beuther02,Shepherd98,Shepherd01,Williams04}. Figure 2 shows a plot of spectral energy distribution.
The best SED fitting gave dust temperature T$_{\rm d}$ of  71.7$\pm$0.4 K, H$_{2}$ gas column
density $N_{\rm H_{2}}$ of (2.7$\pm$1.2)$\times$10$^{24}$ cm$^{-2}$, and
dust emissivity index ($\beta$) of 1.7$\pm$0.4. Derived dust temperature
T$_{\rm d}$ =71.7$\pm$0.4 K is well consistent with SO$_{2}$ rotation
temperature T$^{SO_{2}}_{rot}$ $\sim$ 84$^{+18}_{-15}$ K reported
by \citet{Hauyu13}, indicating that gas is well coupled with dust.

 Gas mass of G192.16 continuum core can be calculated by below formula:
\begin{equation}
M_{\rm H_{2}}  =  \pi R^{2} \cdot m_{\rm H} \cdot \mu \cdot N_{\rm H_{2}},
\end{equation}
where $\mu$ = 2.8 is mean molecular weight \citep{Kauffmann08}, $m_{\rm H}$ is mass of H atom.
$R$ = $\sqrt{ab}D$ is source size, major and minor axes ($a$ and $b$) are obtained from 2D Gaussian fitting toward continuum core (as listed in Table 2). Note that we take major and minor axes ($a$ and $b$) as 1.3$\arcsec$ and 0.7$\arcsec$, which are averaged values of 2D Gaussian fitting results of all continuum sources. At a distance of 1.52 kpc, the core radius is calculated as $R$ = 0.007 pc.
Core mass ($M_{\rm H_{2}}$) is derived to be 10.8$\pm$4.8 M$_{\sun}$, which is consistent with the mass range of 4 M$_{\sun}$ $\leq$ $M_{\rm gas}$+$M_{\rm dust}$ $\leq$ 18 M$_{\sun}$ estimated by \citet{Shiozaki11}.

\begin{figure}[!h]
  \begin{minipage}[t]{0.495\linewidth}
  \centering
   \includegraphics[width=60mm]{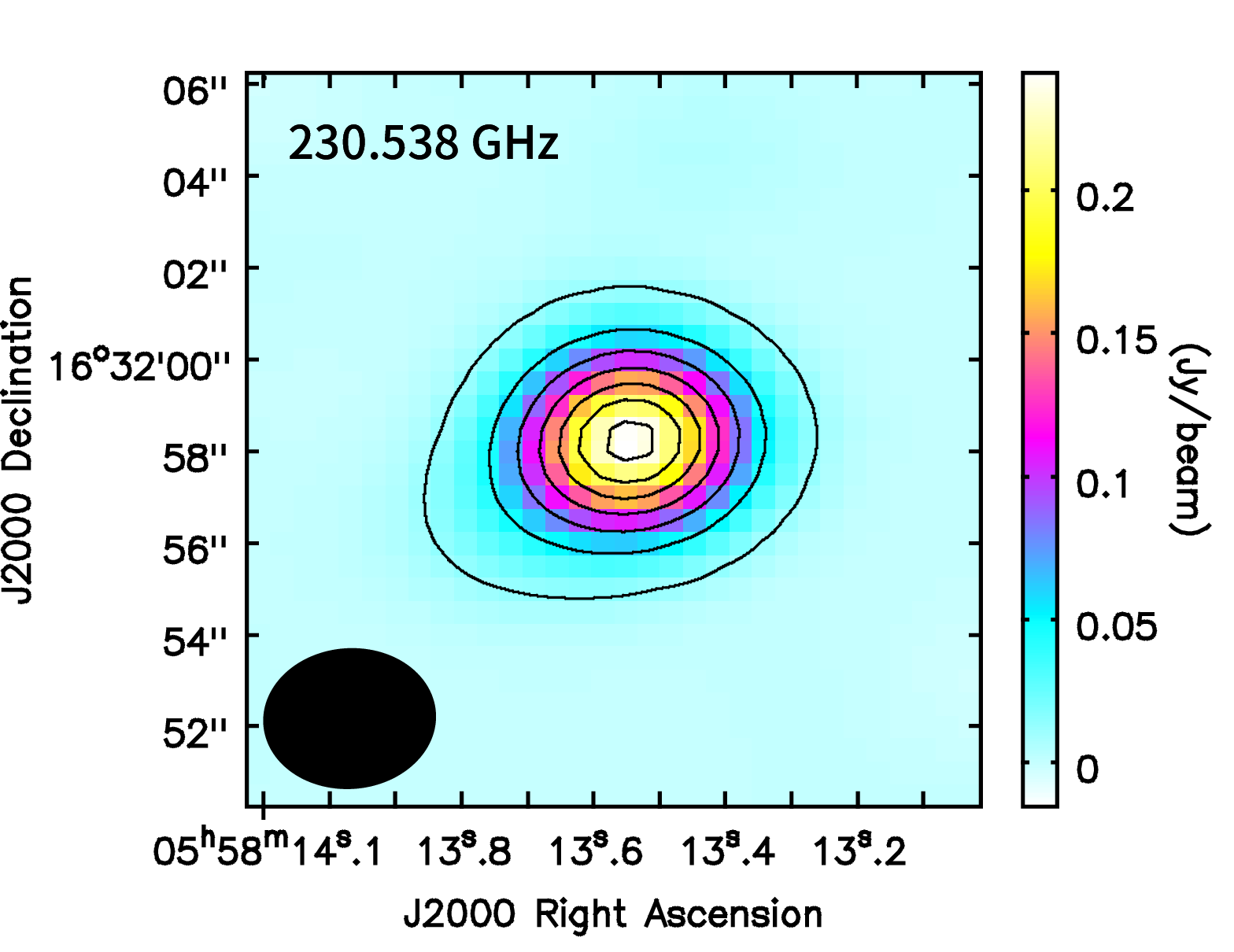}
   \centerline{(a)}
  \end{minipage}%
  \begin{minipage}[t]{0.495\textwidth}
 \centering
  \includegraphics[width=60mm]{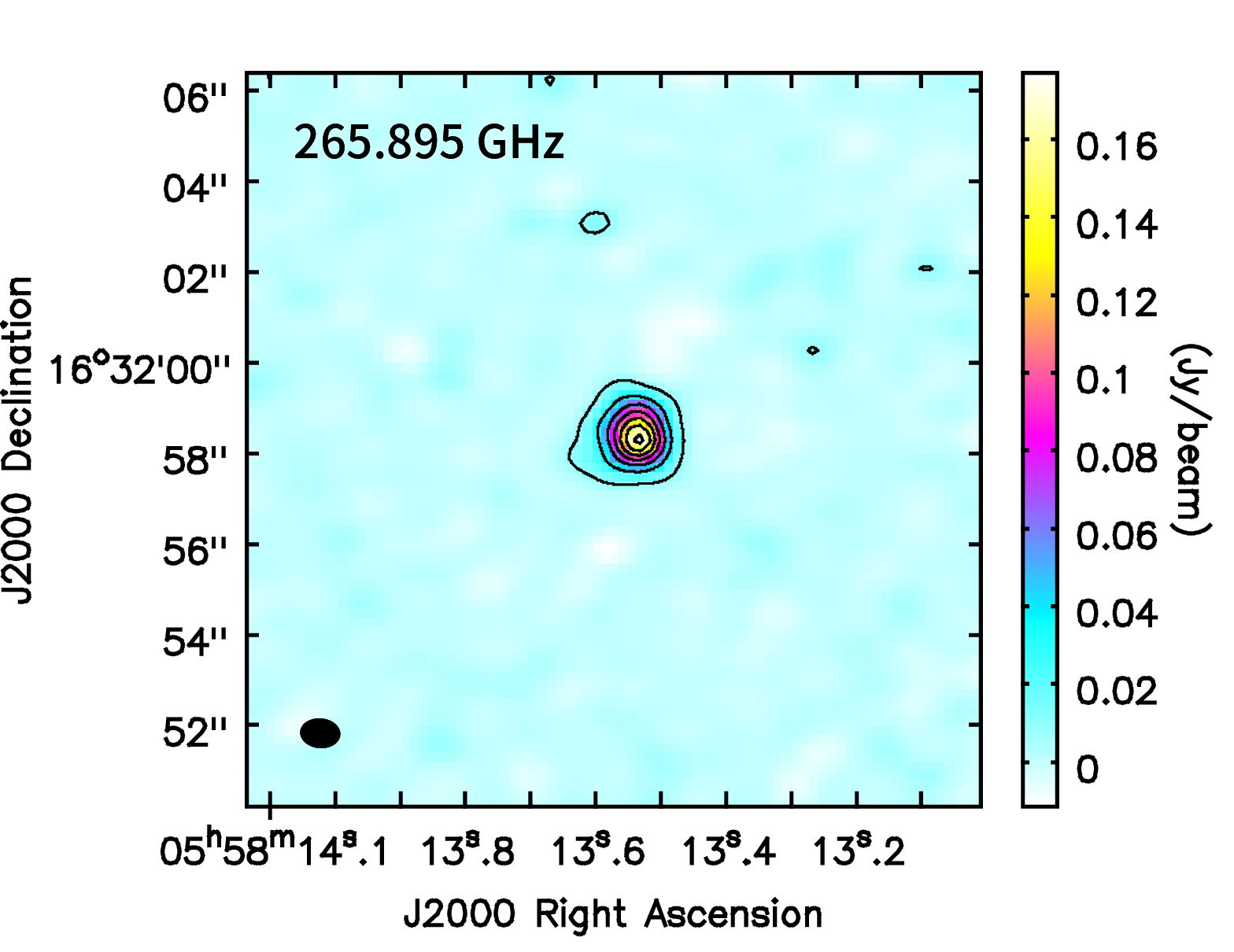}
    \centerline{(b)}
 \end{minipage}%
 \vfill
  \begin{minipage}[t]{1\linewidth}
   \centering
  \includegraphics[width=60mm]{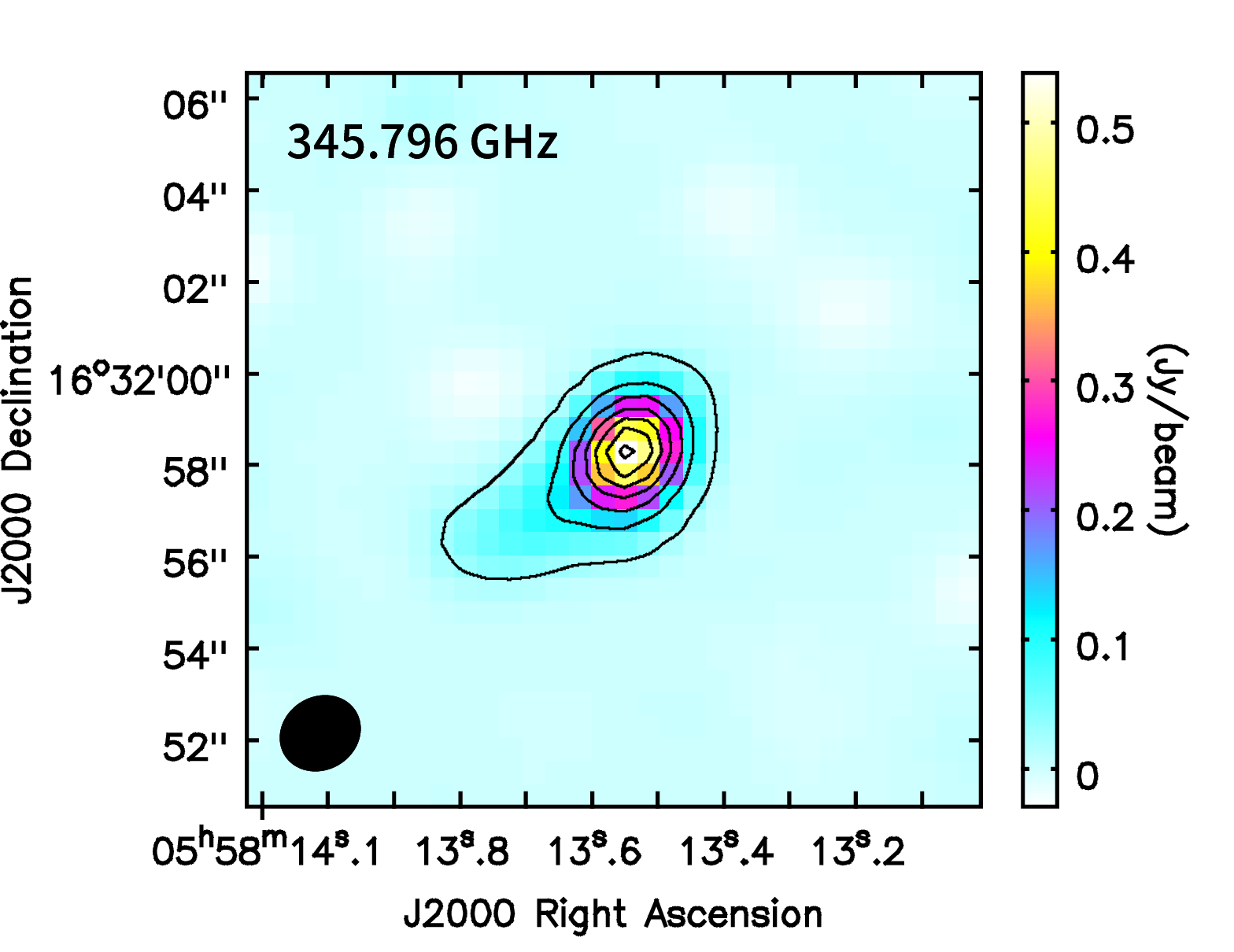}
    \centerline{ (c) }
 \end{minipage}%
 \caption{ Panels (\emph{a}), (\emph{b}), and (\emph{c}) present continuum images of 230.538, 265.895, and 345.896 GHz, respectively. For all panels, the contours are from 5\% to 95\% of of peak values (peak values are shown in Table 2), with a step of 15\%. The synthesized beam size is shown in bottom-left corner of each panel.}
    \label{Fig1}
\end{figure}

\begin{figure}[!h]
\centering
 \includegraphics[width=\textwidth, angle=0]{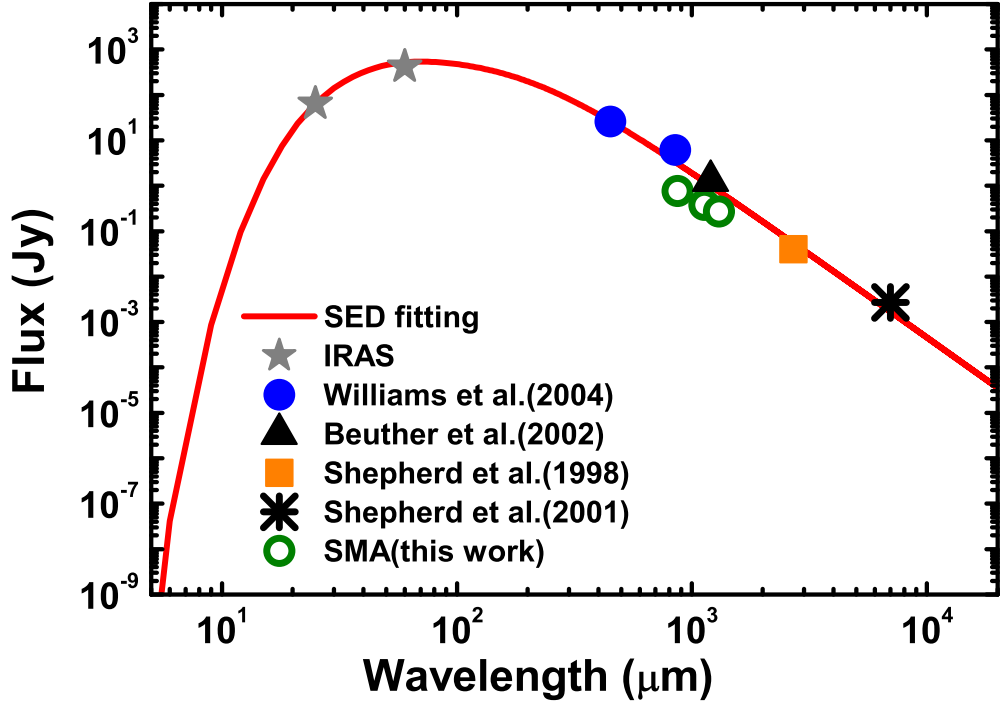}
 \caption{Spectral energy distribution (SED) fitting obtained from our SMA images complemented with literature and archival data at wavelengths ranging from 25 $\mu$m to 7 mm. Red line represents best SED fitting. The data points are shown in gray stars (\emph{IRAS}: 25 $\mu$m and 60 $\mu$m), blue dots \citep{Williams04}, black triangle \citep{Beuther02}, orange square \citep{Shepherd98}, black asterisk \citep{Shepherd01}, and green open circles (this work).}
 \label{Fig2}
\end{figure}

\begin{table}[!h]
\begin{center}
\caption[]{ Parameters of Continuum Images}
\begin{tabular}{ccccc}
  \hline\noalign{\smallskip}
    \hline\noalign{\smallskip}
Frequency    & Deconvolved Size                               & Peak Flux Density  & Total Flux & RMS               \\
GHz          & $a$$\arcsec$$\times$$b$$\arcsec$(P.A.$\degr$)  &Jy beam$^{-1}$      & Jy         & Jy beam$^{-1}$    \\
  \hline\noalign{\smallskip}
230.538 GHz    &1.5$\arcsec$$\times$0.7$\arcsec$(121$\degr$)   &0.239$\pm$0.003 &0.270$\pm$0.003  &0.001 \\
265.895 GHz    &0.9$\arcsec$$\times$0.6$\arcsec$(179$\degr$)   &0.170$\pm$0.004 &0.377$\pm$0.008  &0.002 \\
345.796 GHz    &1.6$\arcsec$$\times$0.8$\arcsec$(146$\degr$)   &0.511$\pm$0.017 &0.769$\pm$0.029  &0.004 \\
  \noalign{\smallskip}\hline
\end{tabular}
\end{center}
\end{table}

\subsection{Molecular lines}
\label{sect:Molecular lines}
Molecular transitions of CO(3-2), CO(2-1), $^{13}$CO(2-1), C$^{18}$O(2-1), HCN(3-2), HCO$^{+}$(3-2), SO(8$_{8}$-7$_{7}$), and SO(6$_{5}$-5$_{4}$) are
detected as shown in Figure 3. In Figure 3, CO(3-2), CO(2-1), $^{13}$CO(2-1),
C$^{18}$O(2-1), HCN(3-2), and HCO$^{+}$(3-2) spectra show double-peaked line
profiles with absorption dips at around $\sim$6 km s$^{-1}$, and the blue-shifted
peaks are stronger than red-shifted ones. While SO(8$_{8}$-7$_{7}$) and SO(6$_{5}$-5$_{4}$) lines show
single peak profiles with LSR velocities at $\sim$ 6 km s$^{-1}$. These double-peaked spectral profiles are so-called ``blue profile'' \citep{Zhou93,Wu03,Wu07}, indicating gas infall in this region.

Various molecular tracers (CO, CN, HCN, H$_{2}$CO, HCO$^{+}$, and etc.) are used for identifying collapse candidates and studying infall \citep{Fuller02,Zapata08,Wu09,Wu14,Liu11a,Liu11b,Liu13a,Liu13b,Pineda12,Qin16,Qiu12}.
Simulations by \citet{Smith12,Smith13} suggested that HCN(3-2) and HCO$^{+}$(3-2) are best ones for
studying gas infall.
From Figure 3, CO(3-2), CO(2-1), $^{13}$CO(2-1) and HCN(3-2) lines reveal much wider line wings, and these line wings may be produced by outflow motions \citep{Shepherd98,Hauyu13}.
Therefore, infall ``profile'' of these lines will be contaminated by outflows. C$^{18}$O(2-1) and HCO$^{+}$(3-2) lines without obvious line
wings will be used for further analyses. Note that observations of C$^{18}$O(2-1) and HCO$^{+}$(3-2) lines have different angular resolution.
We have smoothed higher resolution data (HCO$^{+}$) to lower one (C$^{18}$O).

The integrated intensity maps of C$^{18}$O(2-1) and HCO$^{+}$(3-2) are presented in Figure 4. The red cross of each panel represents the continuum emission peak. For C$^{18}$O(2-1) map, one can see that the gas emission peak is associated with the continuum peak position. However, the HCO$^{+}$(3-2) gas is separated to two components, and one of them is also associated with continuum peak position.

For collapsing cloud, in case that brightness temperature of the background continuum is brighter than excitation temperature
of ``blue profile'' line transition tracing infall motion, the ``blue profile'' line will be becoming
``inverse P-Cygni'' profile.
The modified two-layer model \citep{Myers96,Di01} can fit both  ``blue profile'' and ``inverse P-Cygni'' profile, but also  ``red
profile'' and ``P-Cygni'' profile characterizing of outflows or expansion.
Then the modified two-layer model \citep{Myers96,Di01} is adopted to fit spectral profiles of C$^{18}$O(2-1) and HCO$^{+}$(3-2).
The panels (\emph{a}) and (\emph{b}) of Figure 5 show observed spectra in black and two-layer modelling in red for C$^{18}$O(2-1) and HCO$^{+}$(3-2), respectively.
The two-layer model can be simply described as follow:
\begin{equation}
\Delta T_{\rm B}  =  (J_{f} - J_{cr})[1 - e^{(- \tau_{f})}] + (1 - \Phi)(J_{r} - J_{b})[1-e^{(- \tau_{r} - \tau_{f})}],
\end{equation}
where
\begin{equation}
J_{cr}  =  \Phi J_{c} + (1 - \Phi) J_{r},
\end{equation}
\begin{equation}
\tau_{f}  =  \tau_{0} e^{[\frac{-(V - V_{\rm in} - V_{LSR})^{2}}{2 \sigma^{2}}]},
\end{equation}
\begin{equation}
\tau_{r}  =  \tau_{0} e^{[\frac{-(V + V_{\rm in} - V_{LSR})^{2}}{2 \sigma^{2}}]}.
\end{equation}
The model takes optical depth ($\tau_{0}$), front layer radiation temperature ($J_{\rm f}$), rear layer radiation temperature ($J_{\rm r}$), LSR velocity ($V_{\rm LSR}$),
velocity dispersion ($\sigma$), infall velocity ($V_{\rm in}$), radiation temperatures of the continuum source ($J_{\rm c}$) and fill factor ($\Phi$) into account.
We adopted dust temperature T$_{\rm d}$ = 71.7 K from our SED fitting as radiation temperature ($J_{\rm c}$) of continuum source, and the fill factor ($\Phi$) is fixed to 0.3 during the fitting process. The similar procedure was also used by \citet{Pineda12}.
The $V_{\rm LSR}$ is fixed to 6 km s$^{-1}$, which is derived by SO(6$_{5}$-5$_{4}$) line.
During the fitting process, only $\tau_{0}$, $J_{\rm f}$, $J_{\rm r}$, $\sigma$, and $V_{\rm in}$ are free parameters, the parameter spaces are 0.1$\sim$10 for $\tau_{0}$, 3$\sim$100 K for $J_{\rm f}$ and $J_{\rm r}$, 0.1$\sim$10 km s$^{-1}$ for $V_{\rm in}$. The Levenberg-Marquardt (L-M) algorithm was adopted to search for best solution.

The best fitting gave infall velocities of C$^{18}$O(2-1) and HCO$^{+}$(3-2) spectra are 2.0$\pm$0.2 and 2.8$\pm$0.1 km s$^{-1}$, respectively. The detailed fitting results are presented in Table 3.
We find that $\tau_{0}$, $J_{\rm f}$, and $J_{\rm r}$ are interdependent and very sensitive to initial values. Thus, they can not be determined accurately.
In contrast, $\sigma$ and $V_{\rm in}$ mainly determine the line profile.
Thus, they are much less sensitive to initial values and more reliable.

Assuming that the cloud has a power-law density profile($\rho$$\propto$$r^{1.5}$), the mass enclosed in $r_{0}$ can be calculated by \citep{Liu18}
\begin{equation}
 M  =  \int^{r_{0}}_{0} 4 \pi r^{2} \rho_{0} (\frac{r}{r_{0}})^{-1.5} dr,
\end{equation}
where $r_{0}$ is the outer radius and $\rho_{0}$ is the density at $r_{0}$. Thus, the mass infall rate can be estimated as
\begin{equation}
 M_{\rm in}  = 4 \pi r_{0}^{2} \rho_{0} V_{\rm in} = 1.5 M V_{\rm in}/r_{0}.
\end{equation}
We adopt the mass of 10.8 M$_{\sun}$ from SED fitting for $M$, and averaged source size ($R$) 0.007 pc for $r_{0}$. Thus the mass infall rates of C$^{18}$O(2-1) and HCO$^{+}$(3-2) are estimated to be (4.7$\pm$1.7)$\times$10$^{-3}$ and (6.6$\pm$2.1)$\times$10$^{-3}$ M$_{\sun}$ yr$^{-1}$, respectively. The infall rates are also summarized in Table 3.

\begin{table}[!h]
\bc
\caption[]{The Fitting Results of Two-layer Model}
\setlength{\tabcolsep}{3pt}
 \begin{tabular}{cccccccc}
  \hline\noalign{\smallskip}
    \hline\noalign{\smallskip}
Line    & $\tau_{0}$    &$J_{\rm F}$ &$J_{\rm r}$     &$\sigma$      &$V_{\rm in}$  &$M_{\rm in}$        \\
        &               &(K)         &(K)             &(km s$^{-1}$) &(km s$^{-1}$) &M$_{\sun}$ yr$^{-1}$ \\
  \hline\noalign{\smallskip}
C$^{18}$O(2-1) &0.7$\pm$0.4 &23.8$\pm$2.4 &11.1$\pm$4.7  &0.9$\pm$0.4 &2.0$\pm$0.2 &(4.7$\pm$1.7)$\times$10$^{-3}$  \\
HCO$^{+}$(3-2) &0.4$\pm$0.1 &23.6$\pm$0.3 &5.5$\pm$0.4   &1.4$\pm$0.1 &2.8$\pm$0.1 &(6.6$\pm$2.1)$\times$10$^{-3}$ \\
  \noalign{\smallskip}\hline
\end{tabular}
\ec
\tablecomments{0.86\textwidth}{Optically depth ($\tau_{0}$), front layer radiation temperature ($J_{\rm f}$), rear layer radiation temperature ($J_{\rm r}$), LSR velocity ($V_{\rm LSR}$), velocity dispersion ($\sigma$), infall velocity ($V_{\rm in}$) are free parameters in fitting, while infall rate ($M_{\rm in}$) is calculated by using fitted infall velocity.}
\end{table}

\begin{figure}[!h]
  \begin{minipage}[t]{0.495\linewidth}
  \centering
   \includegraphics[width=60mm]{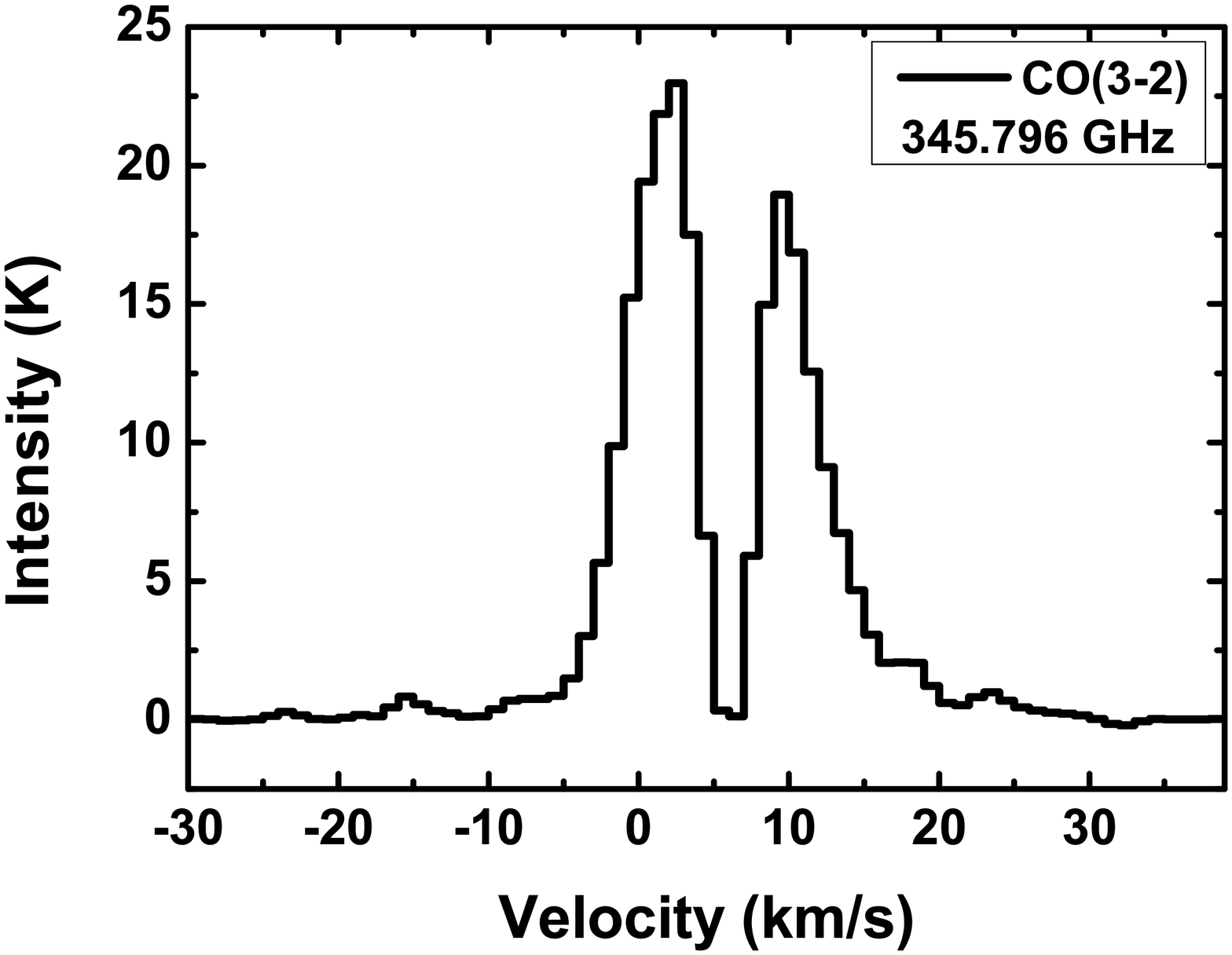}
   \centerline{(a)}
  \end{minipage}%
  \begin{minipage}[t]{0.495\textwidth}
 \centering
  \includegraphics[width=60mm]{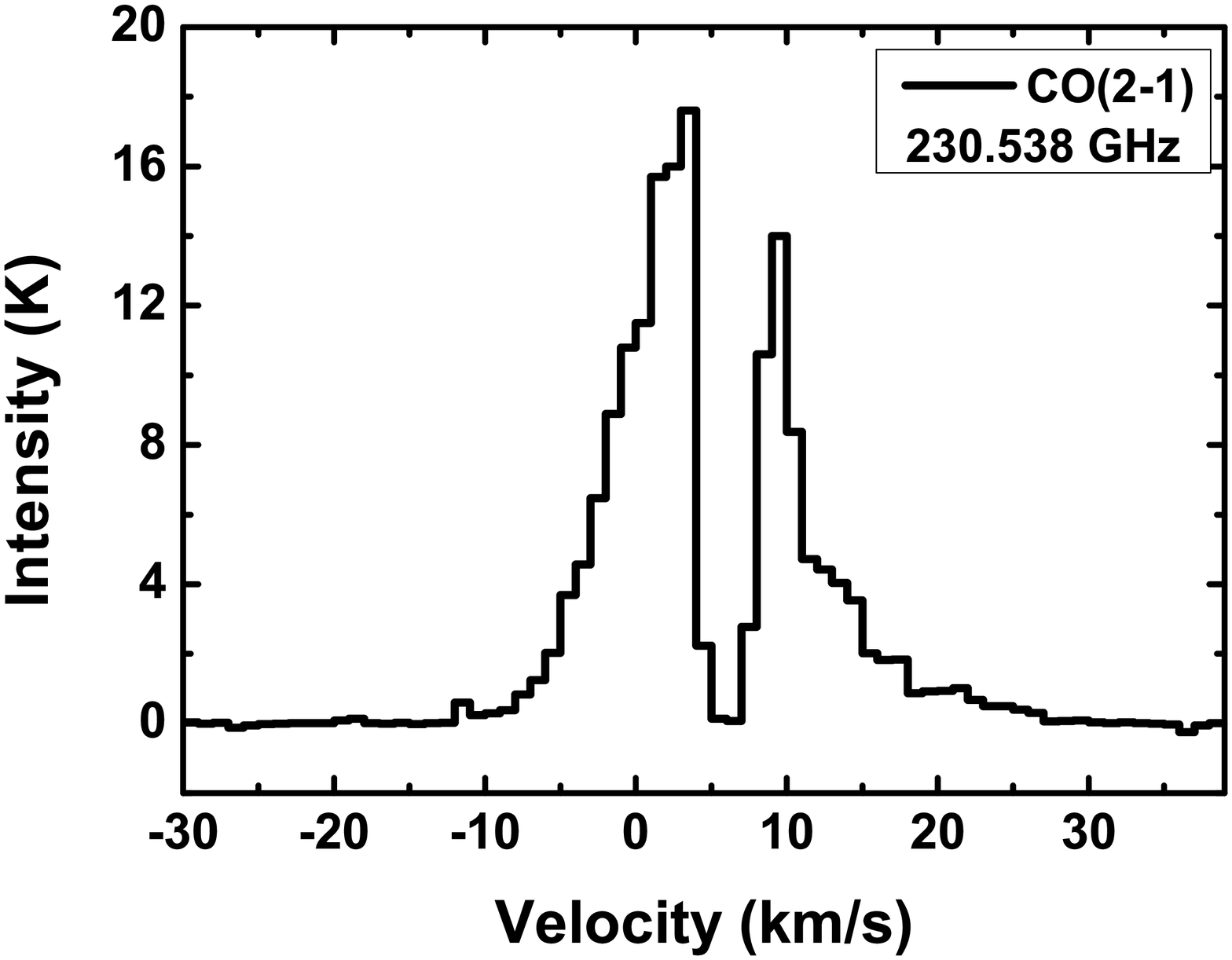}
    \centerline{(b)}
 \end{minipage}%
  \vfill
  \begin{minipage}[t]{0.495\linewidth}
   \centering
  \includegraphics[width=60mm]{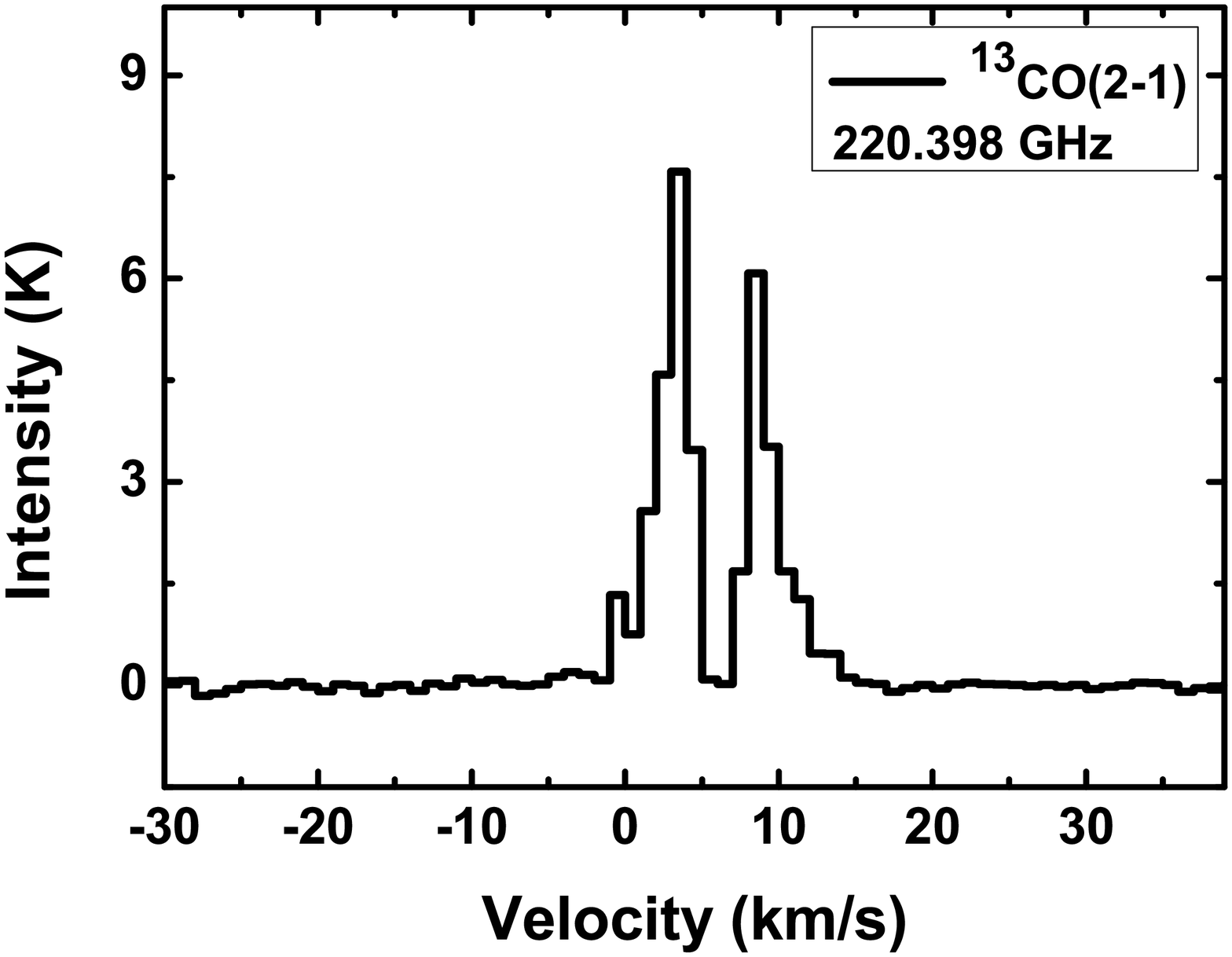}
    \centerline{ (c) }
    \end{minipage}
  \begin{minipage}[t]{0.495\linewidth}
  \centering
  \includegraphics[width=60mm]{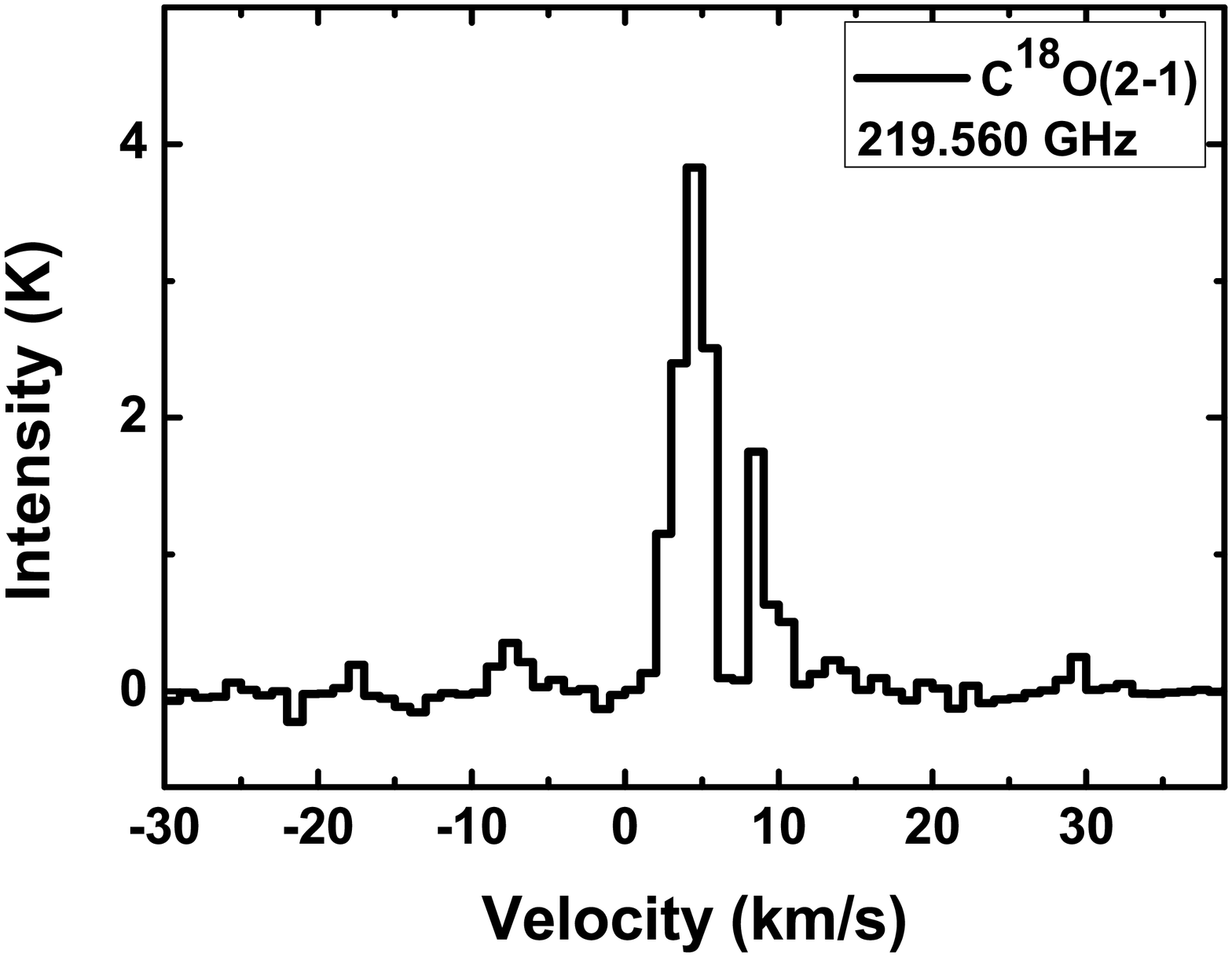}
    \centerline{ (d) }
 \end{minipage}%
 \vfill
   \begin{minipage}[t]{0.495\linewidth}
  \centering
   \includegraphics[width=60mm]{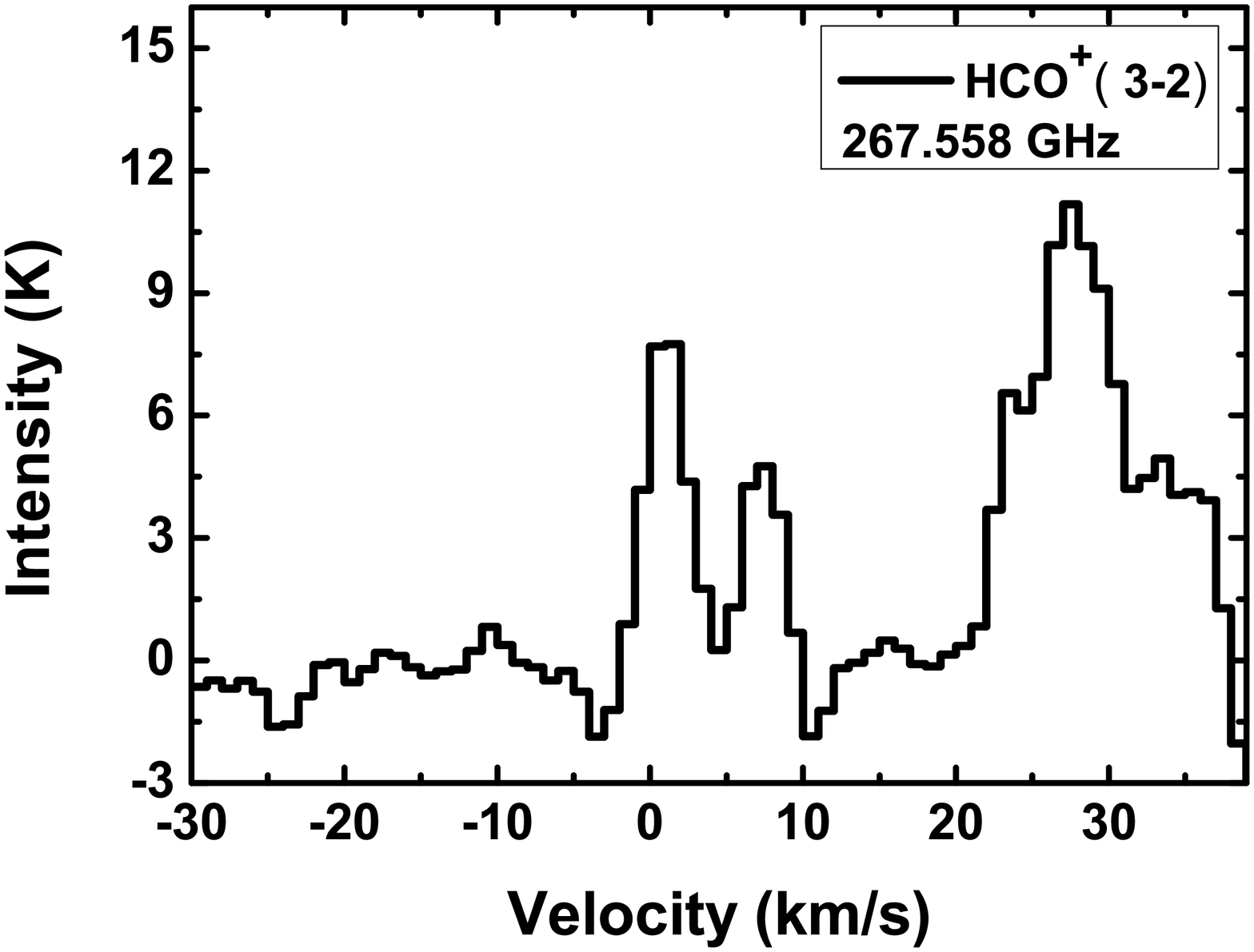}
   \centerline{(e)}
  \end{minipage}%
  \begin{minipage}[t]{0.495\textwidth}
 \centering
  \includegraphics[width=60mm]{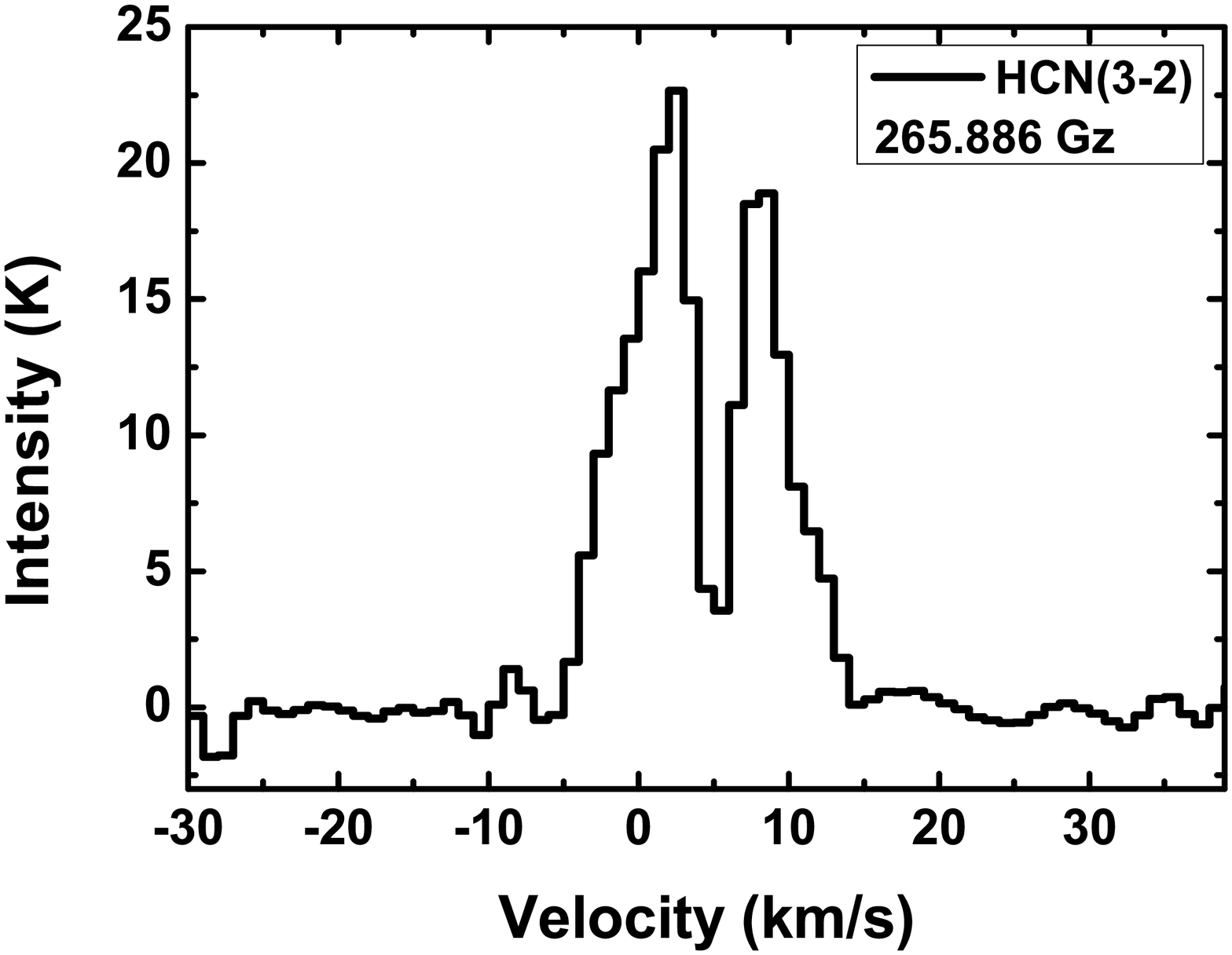}
    \centerline{(f)}
 \end{minipage}%
 \vfill
  \begin{minipage}[t]{0.495\linewidth}
   \centering
  \includegraphics[width=60mm]{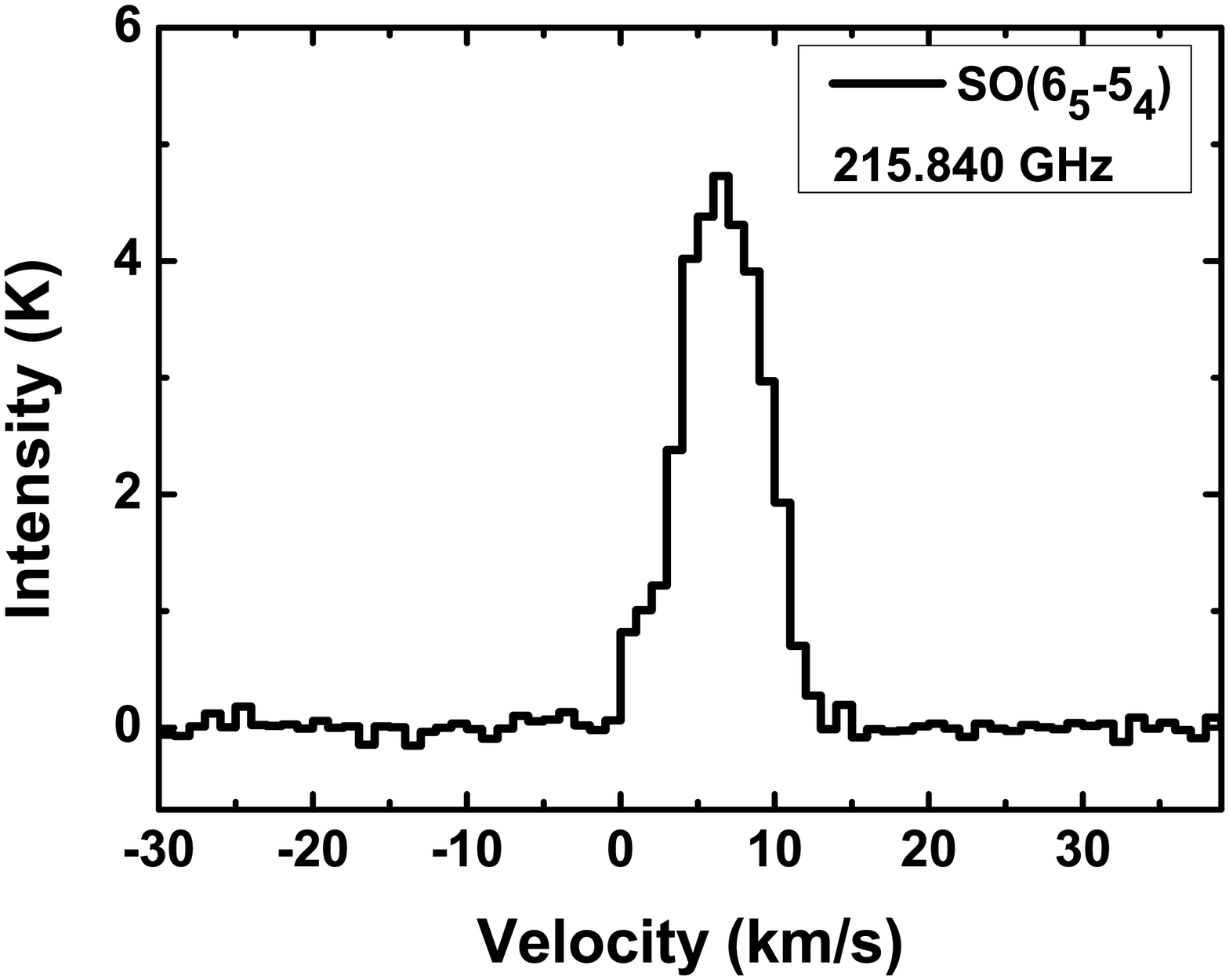}
    \centerline{ (g) }
    \end{minipage}
  \begin{minipage}[t]{0.495\linewidth}
  \centering
  \includegraphics[width=60mm]{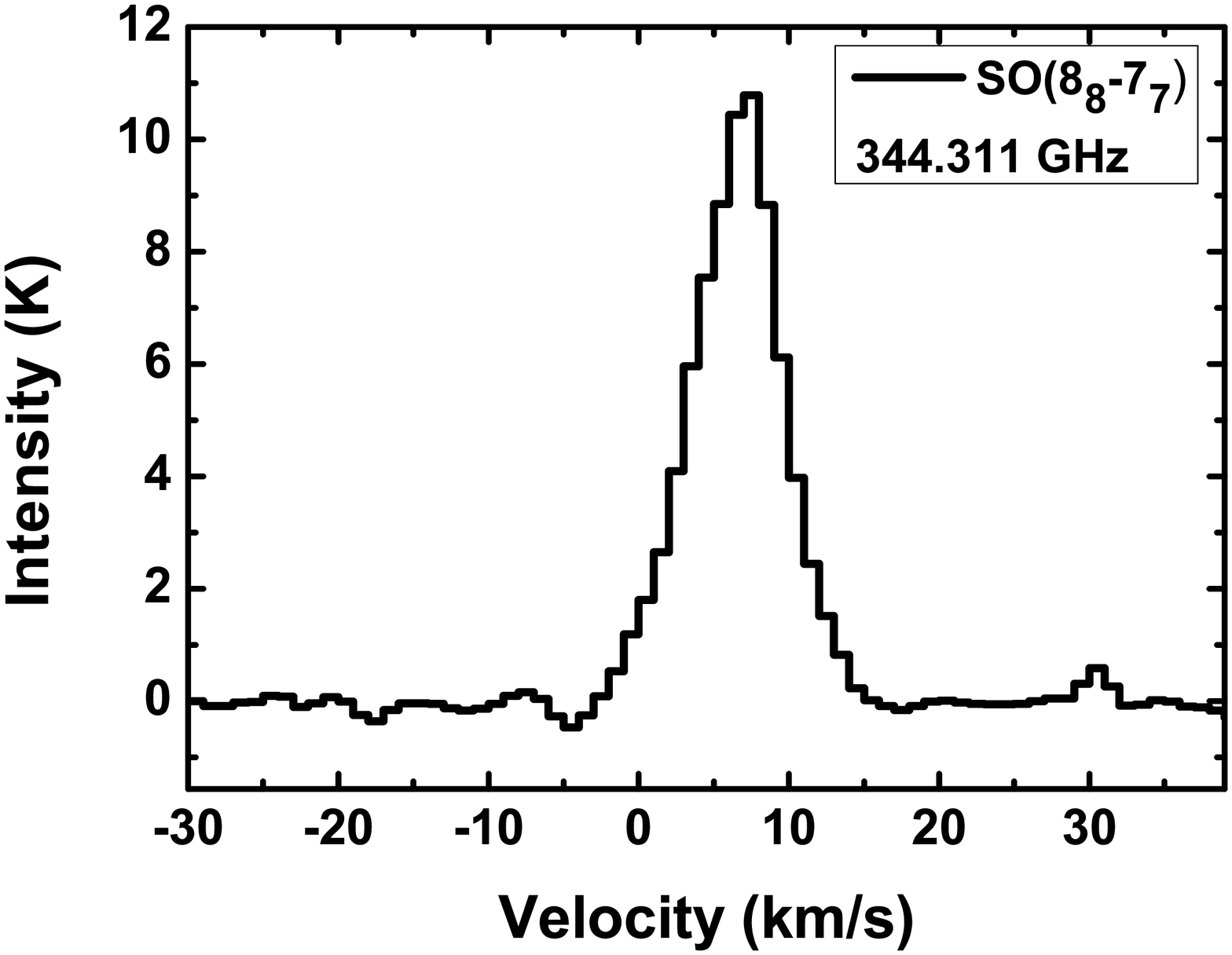}
    \centerline{ (h) }
 \end{minipage}%
 \vfill
 \caption{All spectra are extracted from continuum emission peak position. Line name and rest frequency are shown in upper-right corner of each panel.}
    \label{Fig3}
\end{figure}

\begin{figure}[!h]
  \begin{minipage}[t]{0.495\linewidth}
  \centering
   \includegraphics[width=60mm]{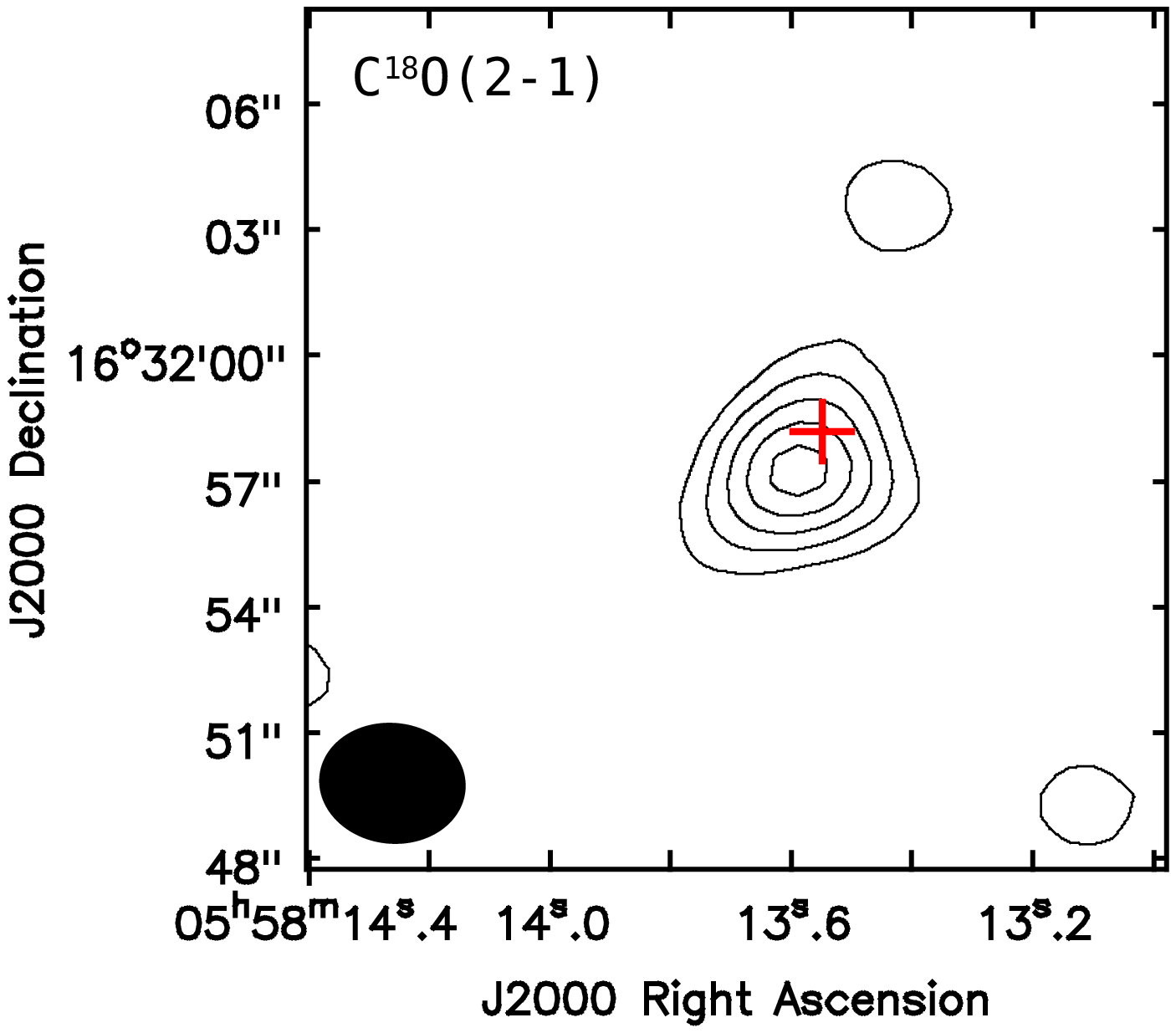}
   \centerline{(a)}
  \end{minipage}%
  \begin{minipage}[t]{0.495\textwidth}
 \centering
  \includegraphics[width=60mm]{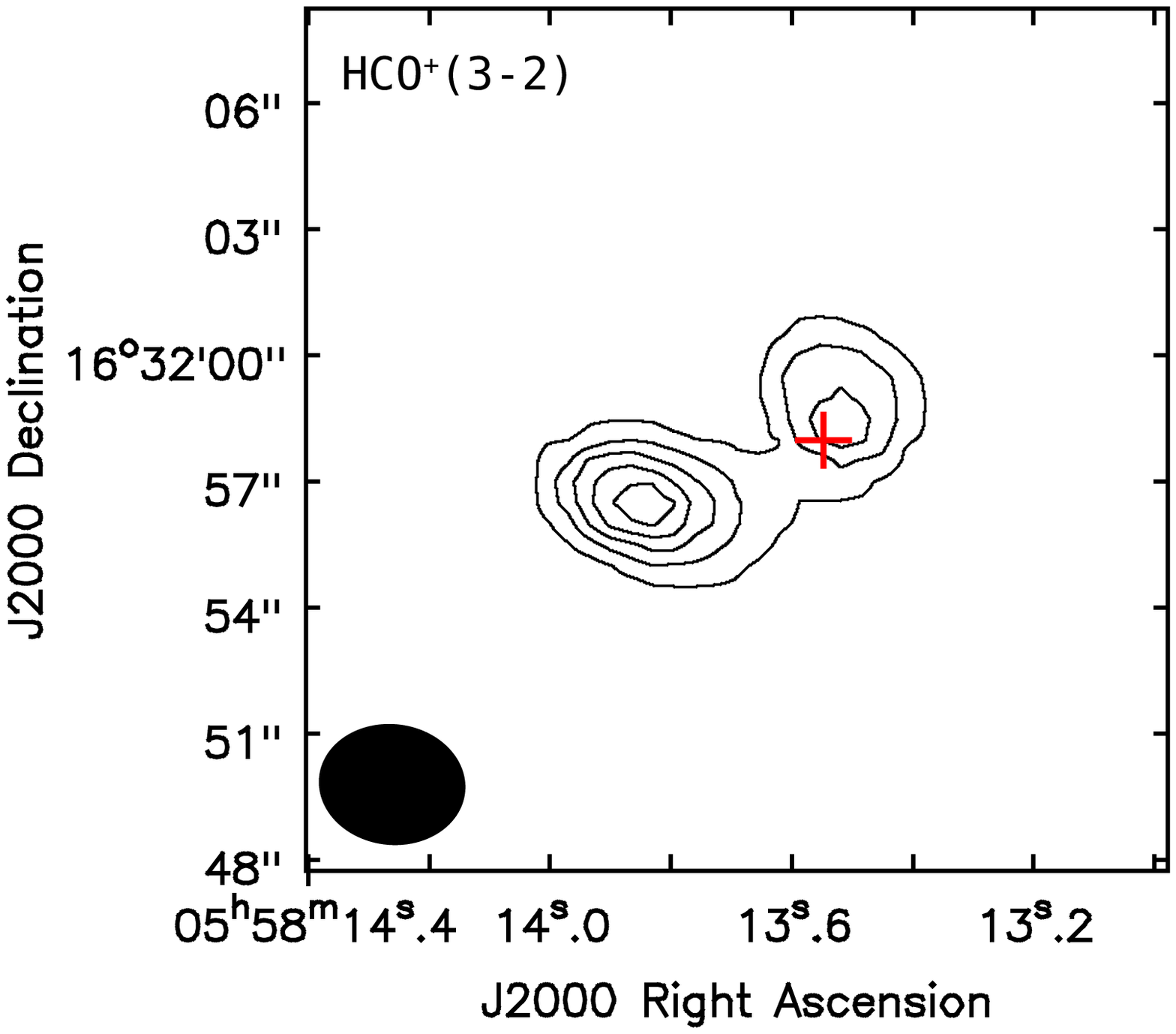}
    \centerline{(b)}
 \end{minipage}%
 \caption{Panels (\emph{a}) and (\emph{b}) present integrated intensity maps of C$^{18}$O(2-1) and HCO$^{+}$(3-2) transitions, repectively. The red cross of each panel is continuum emission peak of G192.16. The contour levels are from 10\% to 90\%, with step of 20\%. The beam sizes of these two observations are shown in left-bottom cornel of each panel. The HCO$^{+}$(3-2) map is smoothed to same resolutions with C$^{18}$O(2-1) map.}
    \label{Fig4}
\end{figure}

\begin{figure}[!h]
  \begin{minipage}[t]{0.495\linewidth}
  \centering
   \includegraphics[width=60mm]{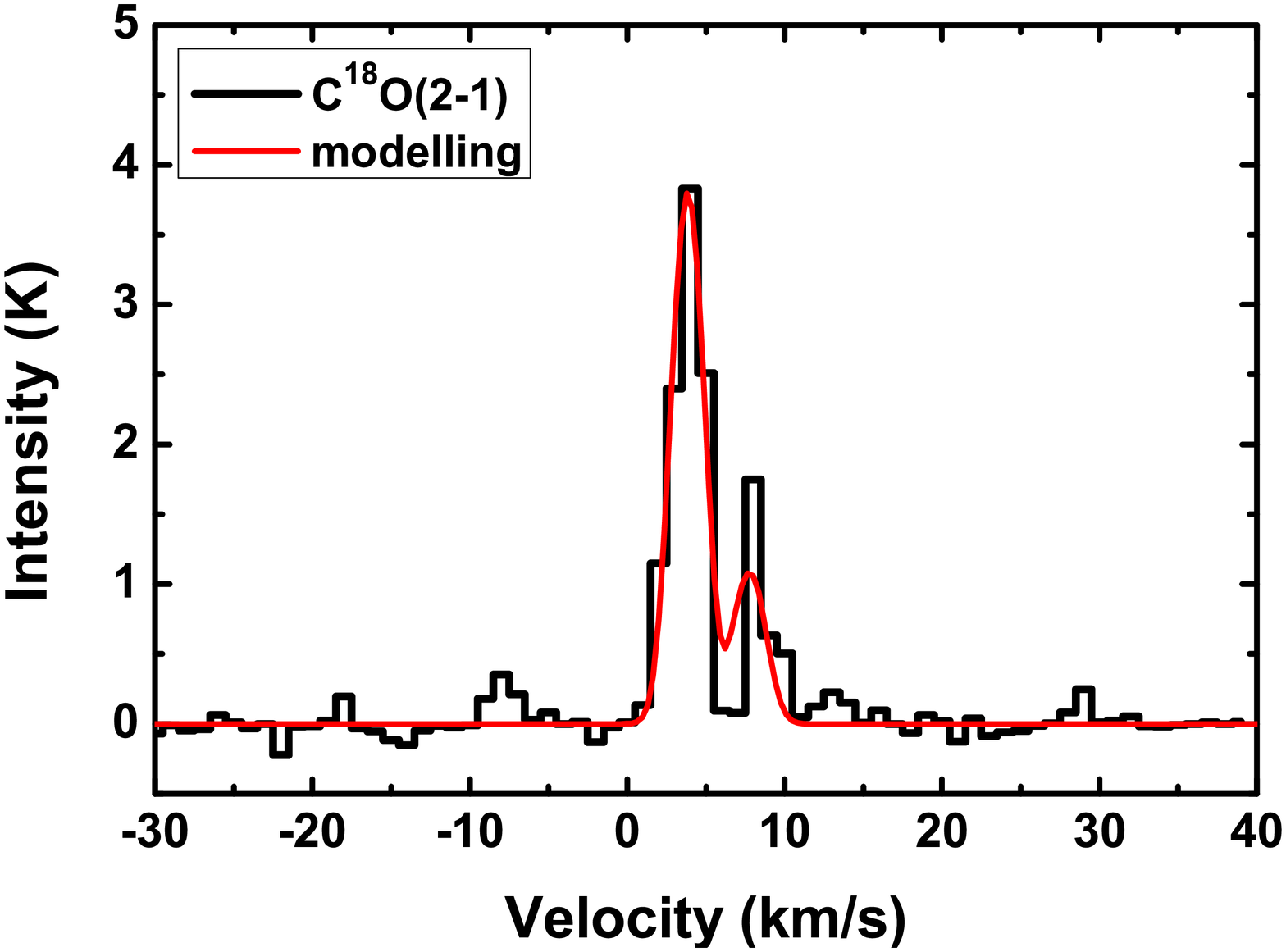}
   \centerline{(a)}
  \end{minipage}%
  \begin{minipage}[t]{0.495\textwidth}
 \centering
  \includegraphics[width=60mm]{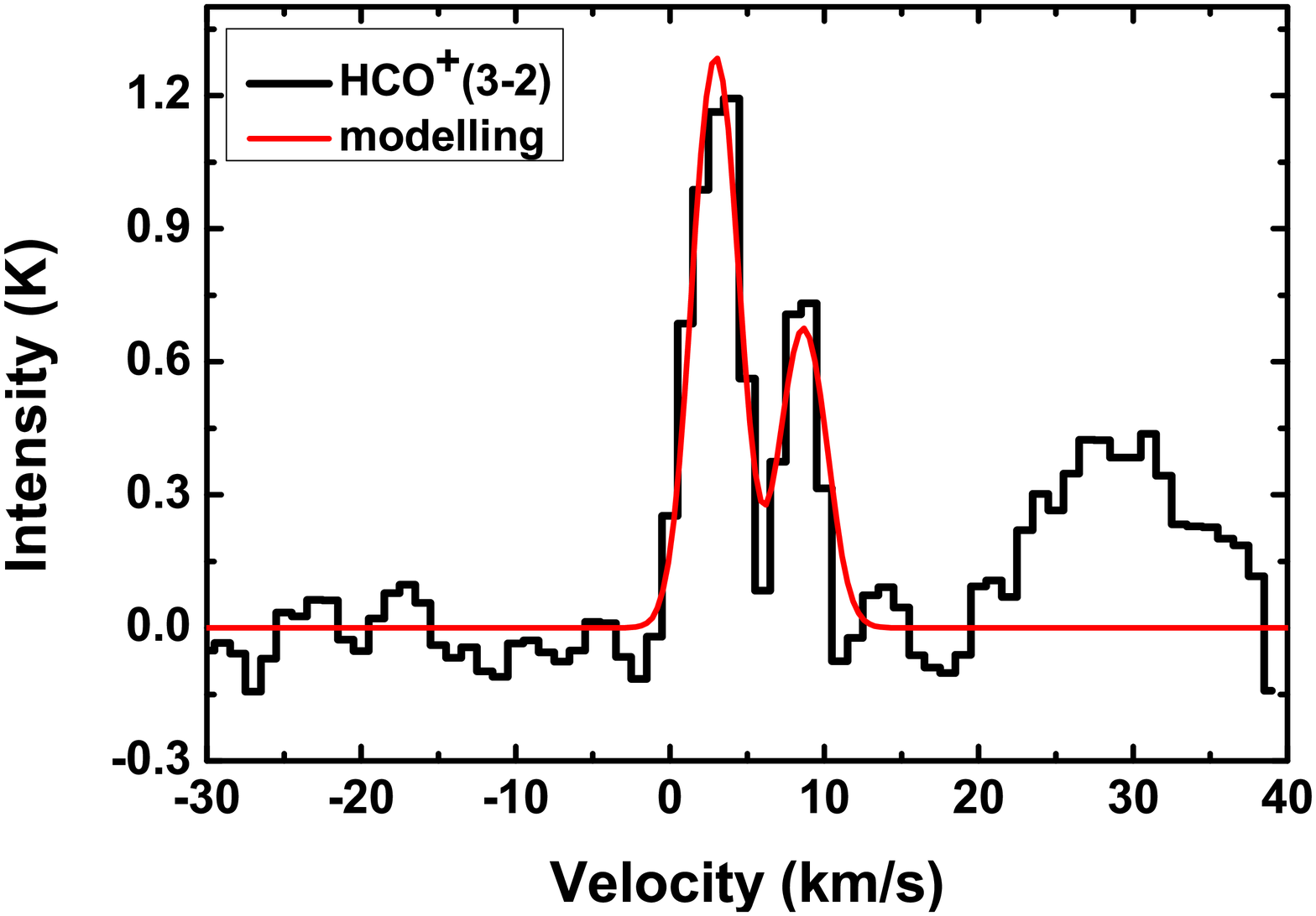}
    \centerline{(b)}
 \end{minipage}%
 \caption{ Panels (\emph{a}) and (\emph{b}) present observed lines (black) and two-layer
modelling (red) of C$^{18}$O(2-1) and HCO$^{+}$(3-2), respectively. The line name is shown in upper-left corner of each panel. The fitting results are listed in Table 3.}
    \label{Fig5}
\end{figure}

\section{Discussion and conclusion}
Rich H$_{2}$O masers \citep{Shepherd04,Imai06,Shiozaki11}, UC
H$_{\rm II}$ region \citep{Hughes93,Shepherd99}, bipolar CO(1-0) outflows
\citep{Shepherd98,Hauyu13}, and an accretion disk \citep{Shepherd01} in
G192.16 have been reported by previous observations, indicating that massive star is forming in this region.

We have identified gas infall in G192.16 region using
C$^{18}$O(2-1) and HCO$^{+}$(3-2) lines, for the first time.
The infall rates derived from these transitions are (4.7$\pm$1.7)$\times10^{-3}$ and (6.6$\pm$2.1)$\times10^{-3}$ M$_{\sun}$ yr$^{-1}$, respectively.
Inflow motions have been reported toward some massive star-forming regions, with mass infall rates ranging from 10$^{-4}$ to
10$^{-2}$ M$_{\sun}$ yr$^{-1}$
\citep{Zhang97,Sandell05,Beltran06,Garay07,Zapata08,Wu09,Wu14,Liu11a,Liu11b,Liu13a,Liu13b,Qin16,Qiu12}.
The derived infall rate toward G192.16 is consistent with those in other
high-mass star formation regions.

In this work, infall rate of C$^{18}$O(2-1) is (4.7$\pm$1.7)$\times10^{-3}$ M$_{\sun}$ yr$^{-1}$.
HCO$^{+}$(3-2) line has higher critical density than that of C$^{18}$O(2-1), and it is generally used for tracing dense and inner parts of molecular clouds.
Infall rate of (6.6$\pm$2.1)$\times$10$^{-3}$ M$_{\sun}$ yr$^{-1}$ is derived
from HCO$^{+}$(3-2), which is larger than that of C$^{18}$O(2-1).
The scenario appears to indicate that infall is faster in inner and denser region than in outer part of the G192.16 core.
This is first time that infall motions have been reported in G192.16 massive core.
The turbulent core model \citep{Mckee03} considers a core having density structure of $\rho$$\propto$$r^{1.5}$, the resulting accretion rate larger than $\sim$10$^{-3}$ M$_{\sun}$ yr$^{-1}$ will be high enough to overcome radiation pressure to form a massive star. In our case, the derived infall rate of $\sim$5$\times$10$^{-3}$ M$_{\sun}$ yr$^{-1}$ by assuming that the dense core have power-law density profile ($\rho$$\propto$$r^{1.5}$). The infall rates of our fitting are consistent of predict of \citet{Mckee03}.
Recent numerical simulations have shown that massive star is formed by disk accretion, the radiation pressure barrier can be easily overcome when an optically thick accretion disk is taken into account\citep{Kuiper10,Kuiper13}. An accretion disk was also reported in G192.16, all these evidences indicate that a massive star is forming in G192.16 core by gas accretion, and high accretion rate is general requirement to form a massive star.
\section{Acknowledgements}
This work has been supported by the
 National Natural Science Foundation of China under grant Nos. 11373026 and
 11433004, and the Joint Research Fund
in Astronomy (U1631237) under cooperative agreement between the
National Natural Science Foundation of China (NSFC) and Chinese
Academy of Sciences (CAS),  by Top Talents Program of Yunnan
Province(2015HA030).

\label{lastpage}


\begin{thebibliography}{99}

\bibitem[Beltr\'{a}n et al.(2006)]{Beltran06}Beltr\'{a}n, M. T., Cesaroni, R., Codella, C., et al. 2006, Nature, 443, 427

\bibitem[Beuther et al.(2002)]{Beuther02}Beuther, H., Schilke, P., \& Menten, K. M. 2002, ApJ, 566, 945

\bibitem[Di Francesco et al.(2001)]{Di01}Di Francesco, J., Myers, P. C., Wilner, D. J., Ohashi, N., \& Mardones, D. 2001, ApJ, 562, 770

\bibitem[Fuller, Williams, \& Sridharan(2002)]{Fuller02}Fuller, G. A., Williams, S. J., \& Sridharan, T. K. 2002, AAS, 200, 7115F

\bibitem[Garay et al.(2007)]{Garay07}Garay, G., Mardones, D., Bronfman, L., et al. 2007, A\&A, 463, 217

\bibitem[Hughes \& MacLeod(1993)]{Hughes93}Hughes, V. A., \& MacLeod, G. C. 1993, AJ, 105, 1495

\bibitem[Imai et al.(2006)]{Imai06}Imai, H., Omodaka, T., Hirota, T., et al. 2006, PASJ, 58, 883

\bibitem[Jiang et al.(2005)]{Jiang05}Jiang, Z., Tamura, M., Fukagawa, M., et al. 2005, Nature, 437, 112

\bibitem[Kauffmann et al.(2008)]{Kauffmann08}Kauffmann, J., Bertoldi, F., Bourke, T. L., Evens, II, N. J., \& Lee, C. W. 2008, A\&A, 487, 993

\bibitem[Kuiper et al.(2010)]{Kuiper10}Kuiper, R., Klahr, H., Beuther, H., \& Henning, T. 2010, ApJ, 722, 1556

\bibitem[Kuiper \& Yorke(2013)]{Kuiper13}Kuiper, R., \& Yorke, H. W. 2013, ApJ, 763, 104

\bibitem[Liu et al.(2011a)]{Liu11a}Liu, T., Wu, Y., Zhang, Q., et al. 2011a, ApJ, 728, 91

\bibitem[Liu et al.(2011b)]{Liu11b}Liu, T., Wu, Y., Liu, S.-Y., et al. 2011b, ApJ, 730, 102

\bibitem[Liu et al.(2013a)]{Liu13a}Liu, T., Wu, Y., Zhang, H., 2013a, ApJ, 776, 29

\bibitem[Liu et al.(2013b)]{Liu13b}Liu, T., Wu, Y., Wu, J., et al., 2013b, MNRAS, 436, 1335

\bibitem[Liu et al.(2013)]{Hauyu13}Liu, H. B., Qiu, K., Zhang, Q., Girart, J. M., \& Ho, P. T. P. 2013, ApJ, 771, 71

\bibitem[Liu et al.(2018)]{Liu18}Liu, T., Li, P. S., Juvela, M., et al. 2018, ApJ, 859, 151

\bibitem[Mckee \& Tan(2003)]{Mckee03}Mckee, Christopher F., \& Tan, Jonathan C. 2003, ApJ, 585, 850

\bibitem[Myers et al.(1996)]{Myers96}Myers, P. C., Mardones, D., Tafalla, M., et al. 1996, ApJ, 465, L133

\bibitem[Patel et al.(2005)]{Patel05}Patel, N. A., Curiel, S., Sridharan, T. K., et al. 2005, Nature, 437, 109

\bibitem[Pineda et al.(2012)]{Pineda12}Pineda, J. E., Maury, A. J., Fuller, G. A., et al. 2012, A\&A, 544, L7

\bibitem[Qin et al.(2008)]{Qin08}Qin, S. -L., Wang, J. -J., Zhao, G., Miller, M., \& Zhao, J. -H. 2008, A\&A, 484, 361

\bibitem[Qin et al.(2016)]{Qin16}Qin, S. -L., Schilke, P., Wu, J., et al. 2016, MNRAS 456, 2681

\bibitem[Qiu et al.(2012)]{Qiu12}Qiu, K., Zhang, Q., Beuther, H., Fallscheer, C. 2012, ApJ, 756, 170

\bibitem[Sandell et al.(2005)]{Sandell05}Sandell, G., Goss, W. M., \& Wright, M. 2005, ApJ, 621, 839

\bibitem[Sault et al.(1995)]{Sault95}Sault, R. J., Teuben, P. J., \& Wright, M. C., 1995, in ASP Conf. Ser. 77, Astronomical Data Analysis Software and systems IV, ed. R. A. Shaw, H. E. Payne, \& J. J. E. Hayes (San Francisco, CA: ASP), 433

\bibitem[Shepherd \& Churchwell(1996)]{Shepherd96}Shepherd, D. S., \& Churchwell, E. 1996, ApJ, 472, 225

\bibitem[Shepherd et al.(1998)]{Shepherd98}Shepherd, D. S., Watson, A. M., Sargent, A. I., \& Churchwell, E. 1998, ApJ, 507, 861

\bibitem[Shepherd \& Kurtz(1999)]{Shepherd99}Shepherd, D. S., \& Kurtz, S. E. 1999, ApJ, 523, 690

\bibitem[Shepherd, Claussen \& Kurtz(2001)]{Shepherd01}Shepherd, D. S., Claussen, M. J., \& Kurtz, S. E. 2001, Science, 292, 1513

\bibitem[Shepherd et al.(2004)]{Shepherd04}Shepherd, D. S., Borders, T., Claussen, M. J., Shirley, Y. \& Kurtz, S. E. 2004, ApJ, 614, 211

\bibitem[Shiozaki et al.(2011)]{Shiozaki11}Shiozaki, S., Imai, H., Tafoya, D., et al. 2011, PASJ, 63, 1219

\bibitem[Smith et al.(2012)]{Smith12}Smith, R. J., Shetty, R., Stutz, A. M., \& Klessen, R. S. 2012, ApJ, 750, 64

\bibitem[Smith et al.(2013)]{Smith13}Smith, R. J., Shetty, R., Beuther, H., et al. 2013, ApJ, 771, 24

\bibitem[Sridharan et al.(2005)]{Sridharan05}Sridharan, T. K., Williams, S. J., \& Fuller, G. A. 2005, ApJ, 631, L73

\bibitem[S\'{a}nchez-Monge et al.(2014)]{Sanche14}S\'{a}nchez-Monge, \'{A}., Beltr\'{a}n, M. T., Cesaroni, R., et al. 2014, A\&A, 569, 11

\bibitem[Williams et al.(2004)]{Williams04}Williams, S. J., Fuller, G. A., \& Sridharan, T. K. 2004, A\&A, 417, 115

\bibitem[Wu \& Evens(2003)]{Wu03}Wu, Jingwen., \& Evans, Neal J., II. 2003, ApJ, 592, L79

\bibitem[Wu et al.(2004)]{Wu04}Wu, Y., Wei, Y., Zhao, M., et al. 2004, A\&A, 426, 503

\bibitem[Wu et al.(2007)]{Wu07}Wu Y. F., Henke S., Xue R., Guan X., \& Miller M. 2007, ApJ, 669, L37

\bibitem[Wu et al.(2009)]{Wu09}Wu, Y. F., Qin, S. -L., Guan, X., et al. 2009, ApJ, 697, L116

\bibitem[Wu et al.(2014)]{Wu14}Wu, Y., Liu, T., \& Qin, S. -L. 2014, ApJ, 791, 123

\bibitem[Zapata et al.(2008)]{Zapata08}Zapata, L. A., Palau, A., Ho, P. T. P., et al. 2008, A\&A 479, L25

\bibitem[Zhang \& Ho(1997)]{Zhang97}Zhang, Q., \& Ho, P. T. P. 1997, ApJ, 488, 241

\bibitem[Zhang et al.(1998)]{Zhang98}Zhang, Q., Hunter, T. R., \& Sridharan, T. K. 1998, ApJ, 505, L151

\bibitem[Zhou et al.(1993)]{Zhou93}Zhou, S., Evans, N. J., II, Koempe, C., \& Walmsley, C. M. 1993, ApJ, 404,
232

\end{thebibliography}
\end{document}